\documentclass[letterpaper, 10 pt, conference]{ieeeconf}  

\IEEEoverridecommandlockouts                              
\usepackage{graphics} 
\usepackage{epsfig} 
\usepackage{amsmath} 
\usepackage{amssymb}  

\usepackage{cite}
\usepackage[refpage,intoc]{nomencl}
\makenomenclature
\newtheorem{prop}{Proposition}
\newtheorem{lemma}{Lemma}
\newtheorem{thm}{Theorem}
\newcommand{\vectornorm}[1]{\lVert#1\rVert}
\newcommand{\G}{\mathcal{G}}
\newcommand{\A}{\mathcal{A}}
\newcommand{\N}{\mathcal{N}}
\newcommand{\Lap}{\mathcal{L}}

\title{\LARGE \bf
On the Stability of Swarm Consensus Under Noisy Control}

\author{Gregory K. Fricke\thanks{Mr. Fricke is a PhD student in Mechanical Engineering and Materials Science at Duke University, Durham, NC 27708, and is a student member of the ASME. gregory.fricke@duke.edu},
~~Bruce Rogers\thanks{Dr. Rogers is Visiting Assistant Professor of Mathematics at Duke University, Durham, NC 27708. bruce@math.duke.edu},
~~Devendra P. Garg\thanks{Dr. Garg is Professor of Mechanical Engineering and Materials Science at Duke University, Durham, NC 27708, and Life Fellow of the ASME. dpgarg@duke.edu}}

\begin{document}
\maketitle
\thispagestyle{empty}
\pagestyle{empty}
\begin{abstract}
Representation of a swarm of independent robotic agents under graph-theoretic constructs allows for more formal analysis of convergence properties. We consider the local and global convergence behavior of an $N$-member swarm of agents in a modified consensus problem wherein the connectivity of agents is governed by probabilistic functions. The addition of a random walk control ensures Lyapunov stability of the swarm consensus. Simulation results are given and planned experiments are described.
\end{abstract}

\section{Introduction}
Control of a swarm of robots may be achieved in many different ways. The seminal work of Reynolds (1987) \cite{Reynolds87flocks} gave the first algorithmically efficient representation of a flock with very simple rules. Reynolds' work was ground-breaking, allowing simulation of the group dynamics involving very large numbers of members. Interestingly, the fundamental work of Braitenberg \cite{Braitenberg84} in the area of behavioral robotics predates Reynolds' research, though that work was focused mostly on individual robots or robot pairs. This idea was further developed by additional research in swarm robotics by Mataric \cite{Mataric92minimizingcomplexity} and Parker \cite{Parker1994}. Early work was quite ad-hoc, but behavioral control continues to find an increasingly rigorous mathematical framework. For variations on this topic, see, e.g., \cite{Ampatzis08, Arkin93communication, Balch98formationcontrol, Chaimowicz07, Holland2005, Lawton2003, Mondada05, Olfati2006, Tanner03stableflocking1, Tanner03stableflocking2}.

Swarming laws may be applied a variety of cooperative-robotics settings. Each of these settings, though, has a primary goal of achieving \emph{agreement} or \emph{consensus}. The specific meaning of agreement in each scenario is different, but in general the implication is that the states of each robot evolve in a coordinated way to achieve some goal. A few specific types of agreement are creation and maintenance of formation \cite{Balch98formationcontrol, Chen2009, Tanner03stableflocking1, Tanner03stableflocking2}, robot rendezvous \cite{Bullo2009}, plume localization \cite{Hayes01swarmodor}, and robot segregation \cite{Kumar2010}.

Formation control may be accomplished under several control regimes. Clearly, a centralized controller could drive robots to a desired configuration under straightforward trajectory control. It is much more difficult and interesting to develop such capability under a distributed control paradigm, allowing greater (if not unlimited) scalability of the swarm. Distributed control of formations falls into the control regimes of potential functions, geometric control, and Graph-based control. Several additional examples of distributed formation can be found in the works of Yun \cite{Yun1997}, Balch \cite{Balch98formationcontrol}, Yamaguchi \cite{Yamaguchi98asympstab}, and Antonelli \cite{Antonelli08}.

Potential field functions are primarily used to balance competing objectives. Generally constructed as inverse-square functions of distance (inspired by physical models of charged particles), potential field functions yield a natural method of balancing the relative importance of such control goals as obstacle avoidance, goal-seeking, and robot following (or avoidance). The use of potential field functions for construction and maintenance of formations has been studied and reported by, among others, Lewis \cite{Lewis1997}, Leonard \cite{Leonard2001}, Olfati-Saber and Murray \cite{Olfati2002a, Olfati2002b, Olfati2006}, Bruemmer \cite{Bruemmer2002}, and Mai \cite{Mai06formationcontrol}.

\section{Graph Representation}
In recent years, several researchers have explored the use of graph-theoretic concepts to provide a more abstract distributed control. Early works in graph-theoretic formation control by Olfati-Saber and Murray \cite{Olfati2002c} and Tanner, Jadbabaie, and Kumar \cite{Tanner03stableflocking1, Tanner03stableflocking2} showed promise.

A brief discussion of graphs is in order. Graph notation varies in the literature; the following notation will be used in this paper. An undirected graph $\G = \G(V,E)$ of order $N$ consists of a set of $N$ nodes or vertices $V_\G = \{1, 2, ... , N\}$ and a set of connections or edges $E_\G \subset \{V_\G \times V_\G\}$. If a pair of nodes is in the set of edges, $(i,j) \in E_\G$, the nodes are \emph{adjacent} and are called \emph{neighbors}, indicated by $i\sim j$ or $ij$. The set of neighbors of a node $i$ is given by $\N_i = \{j \in V_\G~|~j\sim i\}$. Note that in this case, loop edges, e.g., $(i,i) \in E_\G$, are allowed.

When it is clear from the context, the subscript will be dropped, e.g., $V=V_\G$, $E=E_\G$, and $\G(V,E)=\G(V_\G,E_\G)$.%
\nomenclature{$\G$}{A graph, defined by the set of nodes $V$ and edges $E$}%
\nomenclature{$V$}{The set of $N$ nodes in a graph}%
\nomenclature{$V_\G$}{The set of $N$ nodes in graph $\G$}%
\nomenclature{$E$}{The set of edges in a graph}%
\nomenclature{$E_\G$}{The set of edges in graph $\G$}%
\nomenclature{$\N_i$}{The set of neighbors of node $i$ in a graph}%

A \emph{path} from node $i$ to $j$ is an ordered set of nodes, starting at $i$ and ending at $j$, such that each consecutive pair is in $E$. If a path exists from every pair of nodes $(i,j)$ in the graph, the graph is said to be connected. The diameter of the graph is found by finding the minimum path length for each node to every other node; the maximum of these minimum length paths is the graph diameter.

The \emph{degree} of a node is the cardinality of $\N_i$, denoted by $\Delta_i = |\N_i|$. The vector of degrees of the nodes of $\G$ is denoted $\Delta\G = [\Delta_1,...,\Delta_i,...,\Delta_N]^T$. If $\Delta_i=N-1~\forall~i \in V_\G$, i.e., every node is connected to every other node except itself, then graph $\G$ is called complete. A complete graph of order $N$ is also denoted by $K_N$.
\nomenclature{$\Delta_i$}{The degree, or number of neighbors, of node $i$ in a graph}%
\nomenclature{$\Delta\G$}{The vector of degrees of all nodes of graph $\G$}%
\nomenclature{$K_N$}{A \emph{complete} graph with $N$ nodes}%

The complement of graph $\G(V,E)$ is denoted $\overline{\G}$, and is defined on the same node set as $\G$ but with all disconnected nodes in $\G$ are connected, and all connected nodes in $\G$ are disconnected. Formally, given a graph $\G(V,E)$, its complement is $\overline{\G}=\G(V,\overline{E}), \overline{E_\G}=\{(i,j)\in \{V_\G \times V_\G\}~|~(i,j) \notin E_\G, i\neq j\}$.
\nomenclature{$\overline{\G}$}{The complement of graph $\G$}%
\nomenclature{$\overline{E}$}{The complement to set $E$ on the node set $V$}%

The adjacency matrix is an important data structure in the analysis of graphs. The adjacency matrix, for an undirected graph, is a representation of the neighboring relationships among all agents. The adjacency matrix is defined as shown in Eq.~(\ref{eq:adjacency}).\begin{equation}
\label{eq:adjacency}
[\A]_{ij}=\left\{
\begin{aligned}
1\quad&i\sim j\\
0\quad&otherwise
\end{aligned}\right.
\end{equation}%
\nomenclature{$\A$}{The adjacency matrix of a graph}%

The adjacency matrix of undirected graph $\G$ is symmetrix and positive semi-definite. Most importantly though, the adjacency matrix leads to the graph Laplacian, defined as in Eq.~(\ref{eq:graphlaplacian}):
\begin{equation}\label{eq:graphlaplacian}\Lap(\G)=\mathbf{diag}(\Delta\G)-\A.\end{equation}
Note that the degree vector, $\Delta\G$, is equivalent to the vector of row-sums of $\A$. The eigenvalues of the graph Laplacian give information regarding the level of connectivity of the graph.
\nomenclature{$\Lap(\G$)}{The Laplacian of $\G$}%

By its definition, $\Lap(\G)$ will be rank deficient and will have at least one zero eigenvalue with eigenvector $\mathbf{1}_N$. Beyond this, the total number of zero eigenvalues indicates the number of connected components of the graph. For example, $\Lap(\G)$ achieves maximum rank of $N-1$ when the graph is fully connected, thus there is a single zero eigenvalue, indicating that the graph is contained in a single connected group, i.e., the graph is connected. If there are two zero eigenvalues, this indicates that the graph contains two connected subgraphs, i.e., one subgraph with $N-r$ connected nodes and a second connected subgraph with $r$ connected nodes, for any $r\in[1,N-1]$.
\nomenclature{$r$}{The rank of a matrix}%

The values of the eigenvalues also have significance. If the set of eigenvalues, $\Lambda_{\Lap(\G)}$ are re-ordered by magnitude, e.g., $0=\lambda_1\leq\lambda_2\leq...\leq\lambda_N$, then the value of $\lambda_2$ indicates the \emph{algebraic connectivity} of $\G$. As already mentioned, $\lambda_2=0$ if and only if the full graph is not connected. If $\lambda_2\neq0$ however, its value is an indicator of the average degree of the graph. It is also closely related to the graph \emph{diameter} $D_\G$; in fact, the lower bound for a connected graph is $4/(N\times D_\G)$ \cite{Mohar1991}.
\nomenclature{$\Lambda_{\Lap(\G)}$}{The set of eigenvalues of the $\Lap(\G)$}%
\nomenclature{$\lambda_i$}{The $i$-th eigenvalue of a matrix, sorted in increasing real magnitude}%

Closely related to the above definitions is the weighted graph, wherein each edge has an associated weight. Thus, the weighted adjacency matrix is defined as in Eq.~(\ref{eq:weightedadjacency}):
\begin{equation}\label{eq:weightedadjacency}
[\A_w]_{ij}=\left\{
\begin{aligned}
w_{ij}\quad&i\sim j\\
0\quad&otherwise
\end{aligned}\right.
.\end{equation}\nomenclature{$\A_w$}{The matrix representation of a weighted graph}%
 These weights may be representative of many different physical quantities, e.g., they may represent the physical distances between the nodes in Euclidean space, $w_{ij}=\vectornorm{X_i-X_j}_2$, a cost (energy, time, etc.) associated with traversing the edge, etc. The weighted graph Laplacian, $\Lap(\G_w)$, is defined in the same way, i.e., the degree vector is the vector of row-sums of $A_w$.
\nomenclature{$w_{ij}$}{The value of edge $ij$ in a weighted graph}%
\nomenclature{$\G_w$}{A weighted graph}%

Additional information found within the spectra of $\Lap(\G)$, $\A$, $\Lap(\G_w)$, and $\A_w$, are discussed in the 1988 monograph of Cvetkovi\'c \emph{et al} \cite{Cvetkovic1988}, as well as in \cite{Tutte1984, Gross1987, Prisner1995, Bollobas1998, Godsil2001, Beineke2004, Cvetkovic2004}. These books will be used as guides in the development of graph-based control in the proposed research.

Graph-theoretic control of formations is not novel; the reported work of Tanner, Jadbabaie, and Kumar \cite{Tanner03stableflocking1, Tanner03stableflocking2} expounded on this topic. More recently, an extensive set of research has been conducted regarding the controllability of such swarms under graph-theoretic constructs. Mesbahi published several papers \cite{Mesbahi2003, Mesbahi2004, Mesbahi2005} on controllability within a swarm for \emph{agreement}, along with several related works by Hatano, Das, Rahmani, Chen, Kim, and Tan relating to the study and manipulation of the graph Laplacian and its spectra for swarm control \cite{Rahmani2007, Rahmani2009, Chen2009, Kim2005, Tan2008}. Of particular focus in the proposed research is the extension of such graph-theoretic controllability concepts to random graphs along the lines of Hatano and Mesbahi \cite{Hatano2005a, Hatano2005b}. The extension to the dual of controllability, observability, has additionally been explored by Mesbahi and Zelazo \cite{Mesbahi2008, Zelazo2008}.

The key element in the works of Mesbahi is the notion of the state-dependent dynamic graph. The construct of import is a mapping, $g_s$, from the collective system state $x = [x_1, x_2, ... , x_N]^T,\quad x_i \in \mathbb{R}^n, \quad X \in \{x_1 \times x_2 \times ... \times x_N\} \subset \mathbb{R}^{Nn}$, to the graph $\G$.%
\nomenclature{$X$}{The collective system state of size $(Nn\times1)$}%
\nomenclature{$x_i$}{The $n$-dimensional state vector of agent $i$}%
\nomenclature{$N$}{The number of robotic agents in a collective system}%
\nomenclature{$n$}{The dimensionality an individual robot's state-space}%
\nomenclature{$g_s$}{A mapping from the collective state space $X$ to the system graph $\G$}%

The state-dynamic graph mapping $g_s : X \rightarrow \G$ may be defined in many ways, provided $g_s$ is a distance function. A commonly used and intuitive mapping is the Euclidean distance, \begin{equation}\label{eq:gs2norm}g_s = \{ij \in E~|~\vectornorm{x_i - x_j}_2 < \rho,~\rho>0,~i \neq j\}\end{equation} for a specified $\rho$.  The 2-norm has an obvious physical meaning that is directly applicable to real connectivity problems such as wireless networking (where $\rho$ indicates the communication range), or a laser rangefinder (or rangefinder pair) with full $2\pi$ angular coverage or an array of vision sensors (where $\rho$ is the detection range for these sensors). Alternatively, any $p-$norm may be used in a similar fashion, though with somewhat lower physical meaning.
\nomenclature{$\rho$}{Range or distance measurement}%

One such coordinated goal is aggregation. For example, see Equations (\ref{eq:firstorder}) and (\ref{eq:aggregation}) for a swarming law of this type, where $x_i(t)$ is the $n$-dimensional state of agent $i$ at time-step $t$, $u_i(t)$ is the self-determined control of agent $i$ at time-step $t$, $\Psi(t)$ is an objective function possibly independent of the swarm interaction, $\G_i$ is the subgraph of $\G$ including exactly the edges observed by agent $i$, $\Lap(\G_i)$ is the graph Laplacian of this subgraph, and $b_{obj}, b_{agg} \in \mathbb{R}_{\geq0}$ are the relative scaling gains. This model uses simple, $n-$dimensional, first-order dynamics. Additionally, $X_i(t)$ is the ordered, collective state of all agents in $\N_i(t)$.

\begin{equation}\label{eq:firstorder}
x_i(t+1) = x_i(t) + u_i(t)\\
\end{equation}
\begin{equation}\label{eq:aggregation}
u_i(t) = b_{obj}\Psi(t) - b_{agg}\Lap(\G_i)X_i(t)
\end{equation}
\nomenclature{$x_i(t)$}{The $n$-dimensional state of agent $i$ at time-step $t$}%
\nomenclature{$Psi$}{The objective-focused component of control for a swarming agent}%
\nomenclature{$b_{obj}$}{The scalar gain for the objective control component, $Psi$, of a swarming agent}%
\nomenclature{$b_{agg}$}{The scalar gain for the aggregation control law of a swarming agent}%
\nomenclature{$u_i(t)$}{The control applied to agent $i$ at time-step $t$}%

In order for the aggregation to be guaranteed to occur for $b_{obj}=0$, the initial system graph $\G(t=0)$ must be connected, i.e.,\begin{equation*}
X(0) \in \{\mathbb{R}^{Nn} ~|~rank(\Lap(g_s(X(0))))=N-1\}
.\end{equation*}
Alternatively, if $b_{obj}>0$ and the objective function $\Psi$ represents a random walk (e.g., a sample from a uniform distribution on $[-0.5, 0.5]$ taken during time $t$), the probability of successful agreement is 
\begin{equation*}\lim_{t\to\infty}P(agreement)=1,\end{equation*} for finite $N$ and closed space $X$. Note also, though, that for larger values of $N$, motion of clusters within the graph will decrease if the objective function, $\Psi$, is truly a random walk. Thus modifications to the objective function should be made to guarantee continued motion of the swarm. Simple rules include behavioral motion such as that proposed by Reynolds \cite{Reynolds87flocks}, Matari\'c \cite{Mataric92minimizingcomplexity}, or Parker \cite{Parker1994}.

This research work will consider the use of graph-theoretic values, such as the degree vector $\Delta{\G}$ or the eigenvalues of the graph Laplacian $\Lap(\G)$ in manipulating the relative scaling (i.e. importance) of the swarming and objective laws. Before considering this, though, we must first consider the characteristics of the consensus of the swarm.

\section{Proof of Consensus Stability}
Let $D\subset\mathbb{R}^d$ ($d=2$) be compact and convex (and so simply connected).  
For $i=1,\dots N$, $x_i(t)\in  D$ and $x_i(0)\sim\text{UNIF}(D)$.  
Let $\N_i(t)=\{j\in V: \|x_i(t)-x_j(t)\|<R\}$ for some $R>0$.  Here, $R$ is the \emph{vision radius}.  Typically, $R\ll diam(D)$. Note $\N_i(t)\neq\emptyset$ since $i\in\N_i(t)$.

The dynamics of the particles are given by the agreement algorithm plus a noise term
\begin{equation}\label{eq:dynamics}
 x_i(t+1)=|\N_i(t)|^{-1}\sum_{j\in\N_i(t)}x_j(t) + \omega_i^t.
\end{equation}
Because of the noise term, the particles do not converge to consensus.  However, once all the particles are close to one another, they stay close.
\begin{prop}
Taking the dynamics defined by Eq.~\ref{eq:dynamics}, suppose the $\omega_i^t$ are i.i.d. random variables drawn from a distribution that is absolutely continuous with respected to lebesgue measure and supported on some open set $U$ containing the origin.  If $diam(U)<R/2$, then
$$\max_{i.j}\|x_i(t)-x_j(t)\|<R\Rightarrow \max_{i.j}\|x_i(t+1)-x_j(t+1)\|<R.$$
\end{prop}
\begin{proof}
\begin{equation*}\begin{split}
 \max_{i,j}\|x(t+&1)-x_j(t+1)\|\nonumber\\
=&\max_{i,j}\|\frac{1}{N}\sum_{k=1}^Nx_k(t)+\omega_i^t-\frac{1}{N}\sum_{k=1}^Nx_k(t)-\omega_j^t\|\nonumber\\
=&\max_{i,j}\|\omega_i^t-\omega_j^t\|\nonumber\\
\leq&\max_{i,j}(\|\omega_i^t\|+\|\omega_j^t\|)<R\nonumber
\end{split}\end{equation*}
\end{proof}

We now wish to show that eventually the particles must be close to one another.
\begin{lemma}
Taking the dynamics defined by Eq.~\ref{eq:dynamics}, suppose the $\omega_i^t$ are i.i.d. random variables drawn from a distribution that is absolutely continuous with respected to lebesgue measure and supported on some open set $U$ containing the origin, and let $E\subset D$ be open.  Then there is a positive constant $\mu$ and a \emph{deterministic} time $T$ such that for all $i=1,\dots, N$ and $t\geq0$
$$P(x_i(t+T)\in E)>\mu.$$
\end{lemma}
\begin{proof}
Suppose not.  Then there is an open $E_0\subset D$, some agent $\alpha\in V$, and some time $s>0$ such that $P(x_\alpha(t)\in E_0)=0$ for all $t>s$.
Take $r>0$ such that $U$ contains the ball of radius $r$ centered at the origin.  Define the sets $E_{k+1}=\{x\in D: d(x,E_{k})<r\}$, and let 
$$c_\alpha(t)=|\N_\alpha(t)|^{-1}\sum_{j\in\N_\alpha(t)}x_j(t)$$
be the center of agent $\alpha$'s communication group at each time $t$.  For $k>0$, if $c_\alpha(t)\in E_k$, then $P(x_\alpha(t+1)\in E_{k-1})>0$ since $(c_\alpha+U)\cap E_{k-1}$ is non-empty and open and the $\omega_\alpha^t$ are drawn from a measure absolutely continuous with lebesgue measure.  So we must have $c_\alpha(t)\notin E_1$ for all $t>s$; i.e., $P(c_\alpha(t)\in E_1)=0$ for $t>s$.
Now, each $E_k$ is open, $E_k\subset E_{k+1}$ and $D$ is compact.  Thus, there is a finite positive integer $F$ such that $D\subset E_F$.
By finite recursion, $P(c_\alpha(t)\in E_F)=0$ for all $t>s$, a contradiction.
\end{proof}

\begin{thm}
If $diam(U)<R/2$, there is a random time $\tau$ such that
$$P(\max_{i,j}\|x_i(s)-x_j(s)\|<R)=1$$
for all $s>\tau$.
\end{thm}
\begin{proof}
Take any ball of radius $R$ $B_R\subset D$.  By the lemma, there is a positive constant $\mu$ and a deterministic time $T$ such that $P(x_i(T)\in B_R)>\mu$ for all $x_i(0)$.
Then set
$$\mathbf{1}_i(k)=\left\{
\begin{array}{lc}
 1& x_i(kT)\in B_R\\
0& \text{else}.
\end{array}\right.
$$
The $\mathbf{1}_i(k)$ are Bernoulli random variables indexed by $i=1,\dots, N$ and $k=1, 2, \dots$.
Almost surely, there is $k_0\geq1$ such that 
$$\prod_{i=1}^N\mathbf{1}_i(k_0)=1.$$
Thus set $\tau=k_0T$ and apply Proposition 1.
\end{proof}

\subsection{Discussion} 
The theorem takes advantage of the random walk induced by the noise.  In the appropriate product space, there is a unique Lyapunov attractor, and eventually the dynamics will find it.
Significantly, noisy perturbations considered are quite general, so it's possible to apply the theorem if the noise isn't symmetric about the origin.

The convexity of the space $D$ is never used in the proof.  It is needed so that the averages in the agreement algorithm are well-defined.  If we're careful about defining line of sight and which agents participate in the averaging, the assumption can be relaxed to a space that is compact with open interior.  This allows for spaces with obstacles.  If the space is not compact, the agents may wander off in different directions.  This can be addressed by giving the agents a notion of ``center'' of the space so the random walks are recurrent.

\section{Probabilistic Connectivity}
Experience with real systems quickly reveals that sensing and communication ranges are rarely deterministic with range. Intuitively, likelihood of successful communication or detection is a function of range, but is more accurately modeled as a decreasing probability function.

Intuitively, the resolution of most spatial sensors, e.g. cameras or scanning laser rangefinders, is defined in angular space. Thus the spatial resolution decreases with distance; correspondingly, the minimum detectable feature size increases with distance. Based on this observation, we posit that the detection probability decreases inversely with the range. Validation of this assumption in simulation and on our experimental testbed is currently underway, and will be reported in the sequel. Sensing ranges in simulation are commonly represented as a simple circle.  Clearly though, few sensors have full $360^{\circ}$ coverage, thus the orientation of the robot (and its on-board sensor) must be considered in the sensing function as well.

The development of network technologies leading to the World Wide Web have created a robust system for communication.  When networks are wired, communication can generally be assumed to be reliable to within a small amount of timeline jitter. However, with mobile roboitcs when the communication must be performed wirelessly, many factors can contribute to lost communication. The foremost limitation is range, as this dictates the level of coherent radio-frequency power that can be reliably transmitted between two nodes. There are many other factors that contribute to wireless communication reliability, but our model will focus on these. The simplest model of radio-frequency power propagation -- that of a point source and an outgoing spherical wave -- yields that the power decreases with the square of range.

With these observations, we generalize the connectivity to the mathematically more tractable (with respect to continuity of derivatives) decreasing exponential function. The most basic probabilistic connectivity function we will consider is \begin{equation}P((i,j)\in E)=e^{-d_{ij}/\rho}\end{equation} where \begin{equation}d_{ij}=\vectornorm{X_i-X_j}\end{equation} and $\rho$ is a shaping constant. Clearly though, this function is non-zero for $d<\infty$, thus the asymptotic convergence on a non-infinite domain can be easily seen. This is not realistic, as there must be a finite cut-off for these connectivity functions.  Thus we define a maximum range, $R$, such that \begin{equation*}P((i,j)\in E~|~d_{ij}>R)\triangleq0.\end{equation*}

\section{Simulation}
Evolution of swarms under the influence of the averaging and probabilistic edge-formation described above have been investigated in simulation. A practical issue observed during these simulations is that the averaging effect of the consensus rule results in a \emph{decrease} in the large-scale random motion. As all of the agents are subject to random perturbation sampled from the same distribution, we should expect that the average of this noise results in locally constrained motion. However, as seen above, the averaging method described above relies achieving full connectivity, which in turn relies on the ability of the swarm to continue its random motion. Simply increasing the magnitude of the random motion does not solve this problem, as the averaging still occurs and increasing the magnitude too much may overcome the swarm's ability to stay connected.

\begin{figure}[t]
\centering
\includegraphics[width=0.235\textwidth]{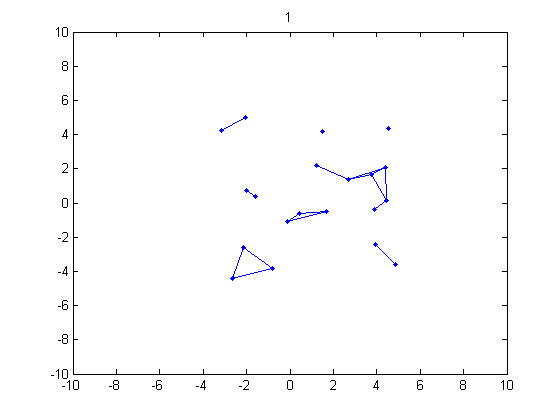}
\includegraphics[width=0.235\textwidth]{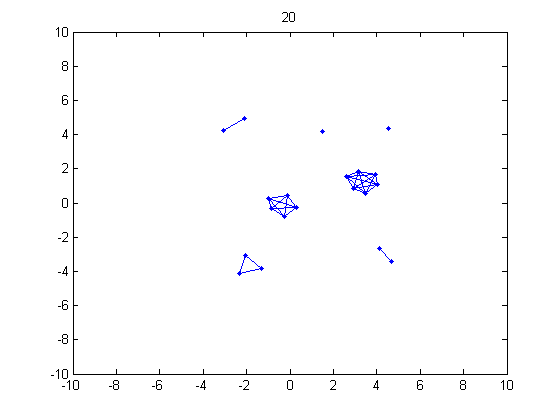}\\

a \qquad\qquad\qquad\quad\qquad\qquad b\\

\includegraphics[width=0.235\textwidth]{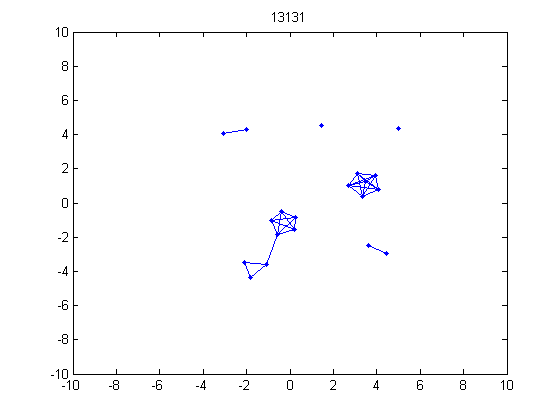}
\includegraphics[width=0.235\textwidth]{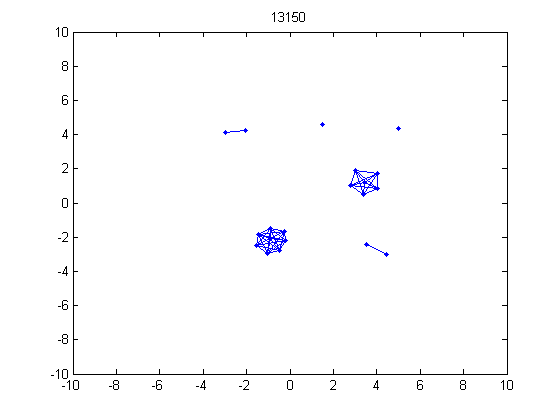}\\

c \qquad\qquad\qquad\quad\qquad\qquad d
\caption{\label{fig:clusters}Cluster formation for 20 agents a) at first iteration (uniform distribution); b) after 20 iterations; c) after 13131 iterations; d) after 13150 iterations.}
\end{figure}

In response, a navigation rule is introduced where in the magnitude of the random perturbation is a function of the individual agent's degree. With this control, individual agents in the interior of a cluster exhibit a larger magnitude of random motion, such that the cluster averaging will not filter the motion. Additionally, given the interior agent's location within the cluster, it is less likely that it's motion in a single iteration will be of large enough magnitude to break all of its connections. Thus greater cluster mobility can be induced without increasing the likelihood of separation. Results from a simulation of this form are given in Fig.~\ref{fig:clusters}.

\section{Planned Experiments}
For verification of the methods described in this paper, an experimental setup with real robots is employed. The arena for robot exploration is an area that is roughly $5m$ x $3.5m$. The agents in this experiment are Create mobile robots from iRobot. A motion capture video system comprised of twelve OptiTrack V100R2 cameras  and the Tracking Tools API from NaturalPoint is utilized to provide an indoor GPS capability. Each Create robot is assigned a unique pattern of reflective markers and a corresponding ``rigid body'' definition in the software, allowing reasonably robust tracking of the agents at 100Hz.

These robots are clearly governed by dynamics that are more complex than single-integrator particles. Experiments will be conducted utilizing real on-board cameras and scanning laser rangefinders (see \cite{Li2011}) to ascertain the limitations of this controller under these dynamics, as well as to understand the rate of convergence under these considerations. Additionally, the probabilistic functions proposed in this paper will be empirically verified. Specifically, the shapes of the sensing functions in two (or more) dimensions must be addressed.

\section{Conclusion}
Swarm agreement under certain probabilistic connectivity limitations has been proven to be asymptotically convergent for bounded spaces. A controller has been developed that conforms to the assumptions in the proof, and simulations verify the convergence under such a controller. Planned experiments will additionally validate these results.

\section{ACKNOWLEDGMENTS}
The authors gratefully acknowledge the support of the Army Research Office under grant number W911NF-08-1-0106.
\addtolength{\textheight}{-19.1 cm}


\end{document}